\documentclass[aps,preprint]{revtex4}
\usepackage{amssymb}

\newcommand{\be}{\begin{equation}}
\newcommand{\ee}{\end{equation}}
\newcommand{\bea}{\begin{eqnarray}}
\newcommand{\eea}{\end{eqnarray}}
\newcommand{\ii}{\'{\i}}
\newcommand{\expq}{e_q}
\newcommand{\lnq}{\ln_q}
\newcommand{\quno}{q-1}
\newcommand{\qunoinv}{\frac{1}{q-1}}
\newcommand{\tr}{{\mathrm{Tr}}}

\begin{document}

\title{
$q$-thermostatistics and the analytical treatment \\
of the ideal Fermi gas
}

\author{S.~Mart\'{\i}nez\thanks{E-mail: martinez@fisica.unlp.edu.ar},
F.~Pennini\thanks{E-mail: pennini@fisica.unlp.edu.ar},
A.~Plastino\thanks{E-mail: plastino@fisica.unlp.edu.ar} and
M.~Portesi\thanks{E-mail: portesi@fisica.unlp.edu.ar}}

\affiliation{
Instituto de F\'{\i}sica La Plata (IFLP),
Universidad Nacional de La Plata (UNLP) and CONICET,
C.C.~727, 1900 La Plata, Argentina }

\vskip 6cm

\begin{abstract}

We discuss relevant aspects of the exact $q$-thermostatistical
treatment for an ideal Fermi system. The grand canonical exact
generalized partition function is given for arbitrary values of
the nonextensivity index $q$, and the ensuing statistics is
derived. Special attention is paid to the mean occupation numbers
of single-particle levels. Limiting instances of interest are
discussed in some detail, namely, the thermodynamic limit,
considering in particular both the high- and low-temperature
regimes, and the approximate results pertaining to the case $q\sim
1$ (the conventional Fermi--Dirac statistics corresponds to $q=1$).
We compare our findings with previous Tsallis' literature.

\hfill

PACS:
05.30.-d, 
05.30.Ch, 
05.30.Fk  

KEYWORDS:
Tsallis' generalized statistics,
Optimized Lagrange multipliers approach,
Thermodynamics,
Ideal Fermi gas

\end{abstract}

\maketitle

\section{Introduction}
\label{SECTintroduction}

Nonextensive
thermostatistics~\cite{tsallisURL,t_csf6,t_bjp29,libro,pp_bjp29,pennini}
constitutes a new paradigm for statistical mechanics. It is based
on Tsallis' nonextensive information measure~\cite{t_jsp52}
\begin{equation}
S_q = k_B\,\frac{1-\sum p_n^{\ q}}{q-1},
\end{equation}
where  $\{p_n\}$ is a set of normalized probabilities and $k_B$
stands for Boltzmann constant ($k_B=1$ hereafter). The real
parameter $q$ is called the index of nonextensivity, the
conventional Boltzmann--Gibbs statistics being recovered in the
limit $q\rightarrow 1$.

The new theory comes in several flavors, though. Within the
literature on Tsallis' thermostatistics, three possible choices
are considered for the evaluation of expectation values in a
nonextensive scenario. As a set of (nonextensive) expectation
values are always regarded as constraints in the associated
$q$-MaxEnt approach~\cite{pp_pla177}, three different generalized
probability distributions ensue. Let $p_n$ ($n=1,\ldots,W$) stand
for the microscopic probability that a system is in the $n$-th
microstate, and consider the (classical) physical quantity $O$
that in the microstate $n$ adopts the value $o_n$. The first
choice~\cite{t_jsp52} for the expectation value of $O$, used by
Tsallis in his seminal paper, was the conventional one: \
$\sum_{n=1}^W p_n \, o_n$. The second choice~\cite{ct_jpa24},
$\sum_{n=1}^W p_n^{\ q} \, o_n$, was regarded as the canonical
definition until quite recently and is the only one that is
guaranteed to yield, always, an analytical solution to the
associated MaxEnt variational problem~\cite{universal}; notice,
however, that the average value of the identity operator is not
equal to one. Elaborate studies of the so-called $q$-Fermi gas
problem, which will constitute the focus of our attention here,
have been performed using this ``Curado--Tsallis
flavor"~\cite{turcos,bdg_pla197,u_pre60,ppp_pa234,tt_pa261}.
Nowadays most authors consider that the third
choice~\cite{pennini,TMP}, usually denoted as the
Tsallis--Mendes--Plastino~(TMP) one,
\begin{equation}
\langle O\rangle_q \equiv \frac{\sum_{n=1}^W p_n^{\ q} \, o_n}
{\sum_{n'=1}^W p_{n'}^{\ q}}
\label{vm3}
\end{equation}
is the most appropriate definition.

We employ the latter choice in order to accommodate the available
a priori information and thus obtain the pertinent probability
distribution via Jaynes' MaxEnt
approach~\cite{jaynes1963,katz1967}, extremizing the $q$-entropy
$S_q$ subject to normalization $\left(\sum_{n=1}^W p_n=1\right)$
and prior knowledge of a set of $M$ nonextensive expectation
values $\{\langle O_j\rangle_q,\ j=1,\ldots,M\}$. As usual the
constrained extremization is accomplished by introducing $M+1$
Lagrange multipliers; in practice two (equivalent) procedures can
be followed to do this. First, the variational procedure followed
by Tsallis--Mendes--Plastino in Ref.~\cite{TMP} gives the Tsallis'
probability distribution in the form
\begin{equation}
p_{n}=\frac{f_{n}^{\ 1/(1-q)}}{\bar{Z}_{q}}
\label{pv}
\end{equation}
where~\cite{TMP}
\begin{equation}
f_{n} = 1-\frac{(1-q)\,\sum_{j=1}^M \lambda_j^{\rm
(TMP)}\left({o_{j}}_{\, n}- \langle O_{j}\rangle_{q}\right)
}{\sum_{n'=1}^W p_{n'}^{\ q}}
\label{fin}
\end{equation}
is the so-called configurational characteristic, and
$\bar{Z}_{q}=\sum_{n} f_{n}^{\ 1/(1-q)}$ represents a ``pseudo"
partition function \ (in the limit $q\rightarrow 1$ it goes to
$Z_1 \, e^{\sum_{j=1}^M \lambda_j \, \langle O_{j}\rangle}$ instead
of $Z_1$). $f_n$ should be positive (otherwise $f_n \equiv 0$) in
order to guarantee that probability $p_n$ be real for arbitrary
$q$ --Tsallis' cutoff condition~\cite{pp_pla193,pp_pla177}--; as a
consequence, the sum in $\bar Z_q$ is restricted to those states
for which $f_n$ is positive.

Notice that the expression obtained for $p_{n}$ following the TMP
recipe is {\it explicitly self-referential}. This fact often leads
to numerical difficulties in concrete applications (see, for
instance, Ref.~\cite{disisto}); more important, it obscures the
underlying physics because the concomitant Lagrange multipliers
loose their traditional physical meaning~\cite{casas}.
Mart\'{\i}nez {\it et al.}~\cite{OLM} devised a way to circumvent
these problems by recourse to the introduction of new, putatively
{\it optimal} Lagrange multipliers (OLM) for the Tsallis'
variational problem. The idea is to extremize the $q$-entropy with
centered mean values (a legitimate alternative procedure) which
entails recasting the constraints in the fashion
\be
\sum_{n=1}^W p_n^{\ q}\left({o_j}_{\, n}-\langle O_j\rangle_q\right)
= 0 \qquad \qquad j=1,\ldots,M
\label{vm3OLM}
\ee
The ensuing microscopic probabilities are formally given by
Eq.~(\ref{pv}), where now
\begin{equation}
f_{n}=1-(1-q)\sum_{j=1}^M \lambda_j\, \left( {o_j}_{\, n}-\langle
O_{j}\rangle_{q}\right)
\label{finueva}
\end{equation}
In this way, the configurational characteristic obtained with the
OLM recipe {\it does not depend explicitly on the set of
probabilities} $\{p_n\}$.

It is obvious that the solution of a constrained extremizing
problem via the celebrated Lagrange method depends exclusively on
the functional form one is dealing with as well as on the
constraints, the Lagrange multipliers being just auxiliary
quantities to be eliminated at the end of the process. As a
consequence, TMP and OLM results should coincide. However their
{\it manipulation} is, in the latter instance, considerably
simpler (notice that the OLM variational
procedure~\cite{OLM,virial,gasideal,temperature,ley0,cuerponegro}
solves {\it directly} for the optimized Lagrange multipliers). Comparing then
the TMP and OLM approaches for a given problem, one realizes that
the resultant probabilities (as well as the pseudo partition
functions) are identical if
\be
\lambda_j = \frac{\lambda_j^{\rm (TMP)}}{\sum_{n=1}^W p_n^{\ q}} =
\bar{Z}_q^{\ q-1} \ \lambda_j^{\rm (TMP)} \qquad \qquad
j=1,\ldots,M
\label{olvida}
\ee
where use has been made of the relation $\sum_n p_n^{\ q}=
\bar{Z}_{q}^{\ 1-q}$~\cite{TMP,OLM} which is valid under the
assumption of the knowledge available a priori.

For the sake of completeness, let us write down~\cite{ley0} the
set of equations that constitute the basic information-theory
relations in Jaynes' version of statistical
mechanics~\cite{jaynes1963,katz1967}:
\begin{eqnarray}
\frac{\partial\ }{\partial \langle O_{j}\rangle_q}
\left(\ln{\bar{Z}_{q}}\right) & = & \lambda_j
\label{termo1}
\\
\frac{\partial\ }{\partial \lambda_j} \left(\ln{Z_{q}}\right)
& = & - \, \langle O_j\rangle_q
\label{termo2}
\end{eqnarray}
for $j=1,\ldots,M$. Here the partition function $Z_q$ is defined
by~\cite{OLM} \ $\ln Z_q \equiv \ln \bar{Z}_q - \sum_{j=1}^M
\lambda_j \, \langle O_j \rangle_q$. While the above OLM equations
involve ordinary logarithms, the analogous TMP
relations~\cite{TMP} employ the so-called $q$-logarithms, $\ln_q
x\equiv(1-x^{1-q})/(q-1)$, thus in the latter case the basic
Jaynes' relations are not recovered and the physical sense becomes
somewhat obscured.

\subsection*{Our goal}

Motivated by the success of the OLM procedure, and in view of some
recent and quite interesting applications of the quantum
distributions
(see, for instance, Ref.~\cite{manuel} for anomalous
behaviors in thermodynamic quantities for ideal Fermi gases
below two dimensions),
we wish to address here with such a technique the
non-interacting Fermi--Dirac gas. The Tsallis generalized
treatment of such system was originally advanced in
Ref.~\cite{turcos} using the Curado--Tsallis flavor.
B\"{u}y\"{u}kk{\i}l{\i}\c{c} {\it et al.} further
investigated~\cite{bdg_pla197} the generalized distribution
functions employing an approximation to deal with nonextensive
quantum statistics called the Factorization Approach~(FA). This
approach, which comes out to give approximate results in the
region $q\sim 1$, is valid for a dilute gas ignoring the
correlations between particles and regarding the states of
different particles as statistically independent. Most succeeding
works on the nonextensive treatment of quantum systems are based
on these approximate generalized distribution functions. The
nonextensive fermion distribution was further analyzed in
Ref.~\cite{ppp_pa234} in the context, again, of the second choice,
but without recourse to that sort of approximations.
It was seen there that the FA faces some difficulties. Other
interesting studies on the subject have been presented by
Ubriaco~\cite{u_pre60} and, quite recently, by Arag\~ao-R\^ego
{\it et al.}~\cite{aslslf_pa317}, employing the third-choice
expectation values along the TMP lines. The latter work focuses
attention upon the thermodynamic limit and provides elegant
analytical results.

The Fermi gas problem is rather cumbersome to treat using the TMP
algorithm. We will show here that the OLM approach allows for an
exact treatment of Fermi distributions in a nonextensive scenario,
in a  simpler way. This, in turn, will make it possible to
re-discuss the nature of the approximation scheme of
B\"{u}y\"{u}kk{\i}l{\i}\c{c} {\it et al.} The paper is organized
as follows: in Sec.~\ref{SECTolm} we sketch the quantum version
of the OLM technique. Sec.~\ref{SECTquantumgas} is devoted
specifically to the grand-canonical description of quantum gases
in a nonextensive framework. Our main results concerning Fermi
systems are developed in Sec.~\ref{SECTfermigas}, where a careful
study of the mean occupancy of discrete single-particle levels is
given. Next we present approximate results for values of $q$ close
to unity, and compare them with previous works on the subject. The
thermodynamic limit is addressed in Sec.~\ref{SECTthermodlimit},
by appealing to integral transform methods that are summarized in
the Appendix. Finally, some conclusions are drawn.

\section{The OLM procedure in quantum language}
\label{SECTolm}

Since we are going to address the ideal Fermi gas, we have to
adapt our nonextensive OLM-Tsallis statistical language to a
quantal environment. Our main tool will be the equilibrium density
operator $\hat{\rho}$, that can be obtained by recourse to the
Lagrange multipliers' method. Within the nonextensive framework
one has to extremize the information
measure~\cite{t_jsp52,t_bjp29,t_csf6}
\begin{equation}
S_{q} [\hat\rho]=\frac{1-\tr\,(\hat{\rho}^{\,q})}{q-1}
\label{entropy}
\end{equation}
subject to the normalization requirement and the assumed a priori
knowledge of the generalized expectation values of, say $M$,
relevant observables, namely
\begin{equation}
\langle\hat{O}_j \rangle_q = \frac{\tr\,(\hat \rho^{\,q} \,
\hat{O}_j)}{\tr\,({\hat \rho}^{\,q})} \qquad \qquad j=1,\ldots,M
\label{gener}
\end{equation}
In the quantum version of the OLM instance the constraints are
recast in the manner
\begin{eqnarray}
\tr\,(\hat{\rho}) & = & 1
\label{normalization}
\\
\tr\left[ \hat{\rho}^{\,q}\left( \hat{O}_j -
\langle\hat{O}_j\rangle_q\right) \right] & = & 0 \qquad \qquad
j=1,\ldots,M
\label{constraints}
\end{eqnarray}
where the $q$-expectation values $\{\langle\hat
O_1\rangle_q,\ldots,\langle\hat O_M\rangle_q\}$ constitute the
external a priori information. Performing the constrained
extremization of Tsallis entropy one obtains~\cite{OLM}
\begin{equation}
\hat{\rho} = \frac{{\hat{f}_q}^{\; 1/(1-q)}}{\bar{Z}_q}
\label{rho}
\end{equation}
where the quantal configurational characteristic has the form
\be
\hat{f}_q = \hat{\openone} - (1-q) \sum_{j=1}^M \,
\lambda_j \; \delta_q\hat{O}_j
\label{charact}
\ee
if the quantity in the right-hand side is positive definite,
otherwise $\hat f_q=0$ --cutoff condition~\cite{pp_pla193,TMP}--.
Here $\{\lambda_1,\ldots,\lambda_M\}$ stands for the set of
optimal Lagrange multipliers, and we have defined for brevity the
generalized deviation as \ $\delta_q\hat{O}\equiv
\hat{O}-\langle\hat{O}\rangle_q$. The normalizing factor in
Eq.~(\ref{rho}) corresponds to the OLM generalized partition
function which is given, in analogy with the classical situation,
by~\cite{OLM,virial,gasideal,temperature,ley0,cuerponegro}
\be
\bar{Z}_q = \tr \, \left( {\hat{f}_q}^{\; 1/(1-q)}\right) = \tr \,
\left[ \, \expq \left( - \sum_{j=1}^M \lambda_j~\delta_q\hat{O}_j
\right) \right]
\label{Zqbar}
\ee
where the trace evaluation is to be performed with due caution in
order to account for Tsallis' cutoff, and
\be
\expq (x) \equiv [1+(1-q)x]^{1/(1-q)}
\label{expq}
\ee
is a generalization of the exponential function, which is
recovered when $q\rightarrow 1$. Let us remark that the density
operator \`a la OLM is not self-referential.

It is to be pointed out that within the TMP framework one obtains
from the normalization condition on the equilibrium density
operator $\hat\rho$ the following relation that the OLM approach
inherits~\cite{TMP,OLM,virial,gasideal,temperature,ley0,cuerponegro},
namely,
\begin{equation}
\tr\,\left[\hat{f}_q^{\; 1/(1-q)}\right] =
\tr\,\left[\hat{f}_q^{\; q/(1-q)}\right]
\label{relac1}
\end{equation}
Making use of this relation, one can obtain the value of the
extremized $q$-entropy as
\begin{equation}
S_{q} =  \lnq\left(\bar{Z}_q\right) \label{S2}
\end{equation}

For the sake of completeness, we can write down the generalized
mean value of a quantum operator $\hat O$ in terms of the quantal
configurational characteristic as
\be
\langle\hat{O} \rangle_q = \frac{\tr\,\left[{\hat
f_q}^{\; q/(1-q)} \, \hat{O}\right]} {\tr\,\left[{\hat
f_q}^{\; q/(1-q)}\right]}
\label{generf}
\ee

We recapitulate in the Appendix how to employ a quite useful
method for calculating the generalized partition function and
expectation values of relevant operators, by recourse to suitable
integral representations. The procedure, which is based on the
definition of the Euler gamma function, has been used in the
literature by many
authors~\cite{tsallis1994,prato,lenzi,aslslf_pa317,se_0302094,virial,gasideal,cuerponegro}
as it enables to express $q$-generalized quantities in terms of
the conventional ($q=1$) ones, thus providing an alternative
analytic approach.

\section{Quantum gas in a generalized grand canonical ensemble}
\label{SECTquantumgas}

It is our aim here that of developing formally the statistical
description of quantum systems, particularly fermions, in a
generalized framework. For this purpose, we will make use of the
OLM-Tsallis version of nonextensive statistics. The Hamiltonian of
the system is assumed to be of the form
\[
\hat{H} =
\sum_k \epsilon_k \, \hat{n}_k
\]
while the number operator exhibits the appearance $\hat N = \sum_k
\hat n_k$, where $\epsilon_k$ and $\hat n_k$ denote, respectively,
the energy and occupation number operator of the $k$-th
single-particle (s.p.) level for a discrete-energy spectrum,
with $\epsilon_1<\epsilon_2<\ldots$

Following a grand-canonical-ensemble description, the
configurational characteristic is given by
\be
\hat f_q = \hat{\openone} - (1-q)
\beta\,(\hat H-U_q) - (1-q) \alpha\,(\hat N-N_q)
\label{charactGC}
\ee
where the Lagrange multipliers $\beta$ and $\alpha\equiv -
\beta\mu$ are related to the temperature and chemical potential
$\mu$ of the system, respectively, and we have designated the
generalized mean values of $\hat H$ and $\hat N$ by $U_q$ and
$N_q$, respectively. The OLM generalized grand partition function
for this ideal quantum gas is obtained by inserting the last
expression into Eq.~(\ref{Zqbar}).

In order to simplify the notation we now define
$\epsilon_k^*\equiv\epsilon_k-\mu$ and the Legendre transform
(``free energy") $\hat H^*\equiv\hat H-\mu\hat N = \sum_k
\epsilon_k^* \, \hat n_k$. The quantities whose mean value is
assumedly known, i.e.\ the internal energy and the particle
number, are then combined in the fashion
\be
\langle \hat{H}^* \rangle_q = U_q-\mu N_q \equiv
U_q^*
\ee
This notation allows one to treat the grand canonical ensemble as
if it were the canonical one, but with grand canonical traces. In
terms of the new (``star") quantities, the configurational
characteristic reads
\be
\hat{f}_q =
\hat{\openone} - (1-q) \beta \, (\hat{H}^*- U_q^*) \equiv
\frac{\beta}{\beta_q} \, \hat{g}_q
\label{charactg}
\ee
where
\be
\beta_q = \frac{\beta}{1+(1-q)\beta U_q^*}
\ee
and
\be
\hat{g}_q = \hat{\openone}-(1-q)\beta_q \hat H^*
\ee
are auxiliary quantities.
(It will be seen below that $\beta_q$ corresponds to the inverse
temperature, while $\hat{g}_q$ to the configurational
characteristic, in a Curado--Tsallis
treatment~\cite{turcos,bdg_pla197,u_pre60,ppp_pa234,tt_pa261}.)

Our partition function can therefore be written as
\bea
\bar{Z}_q & = &
\left(\frac{\beta}{\beta_q}\right)^{1/(1-q)}
\tr\,\left( \hat{g}_q^{\; 1/(1-q)} \right)
\label{Zqbarg} \\
& = & \expq(\beta U_q^*) \ \tr\,\left[\expq(-\beta_q\hat
H^*)\right] \nonumber
\eea
where the last expression resembles the form of this function in
conventional statistics. Once again, the trace entails the cutoff
restriction which now reads: $(\beta/\beta_q)\,\hat g_q$ {\it
should be positive definite}. {}From Eqs.~(\ref{generf})
and~(\ref{charactg}), the scalar quantity $U_q^*$ becomes
\be
U_q^* = \frac{\tr\,\left( \hat{g}_q^{\; q/(1-q)}\,\hat H^*\right)}
{\tr\, \left(\hat{g}_q^{\; q/(1-q)}\right)}
\label{generg}
\ee
Similar expressions hold for $U_q$ and $N_q$ separately, employing
the corresponding operator inside the upper trace. A comparison of
these generalized mean values with the concomitant Curado--Tsallis
results~\cite{turcos,bdg_pla197,u_pre60,ppp_pa234,tt_pa261} allows
the identification of the auxiliary quantity $\beta_q$ with the
inverse temperature defined in the unnormalized context.

The generalized heat capacity is defined as
\be
{C_V}_q \equiv \left. \frac{\partial U_q}{\partial T}\right|_{N_q,V} =
\left. - 
\beta^2 \, \frac{\partial U_q}{\partial\beta}\right|_{N_q,V}
\label{Cq}
\ee
After some algebra we can give it formally in the following fashion
\be
{C_V}_q = 
\frac{1}{1-q} \,
\frac{q\beta_q \, [ \langle\hat g_q^{-1}\,\delta_q\hat H\rangle_q -
\langle\hat g_q^{-1}\,\delta_q\hat H\,\delta_q\hat N\rangle_q \,
\langle\hat g_q^{-1}\,\delta_q\hat N\rangle_q \, / \,
\langle\hat g_q^{-1}\,(\delta_q\hat N)^2\rangle_q]
}
{1-q\beta_q \, [\langle\hat g_q^{-1}\,\delta_q\hat H\rangle_q -
\langle\hat g_q^{-1}\,\delta_q\hat H\,\delta_q\hat N\rangle_q \,
\langle\hat g_q^{-1}\,\delta_q\hat N\rangle_q \, / \,
\langle\hat g_q^{-1}\,(\delta_q\hat N)^2\rangle_q]}
\ee

\section{Ideal Fermi gas results}
\label{SECTfermigas}

In this section we present our fundamental results related to the
nonextensive description of fermion systems, emphasizing the
consequences of dealing with a {\it discrete} single-particle
energy spectrum. We start by reminding that in the conventional
statistical treatment of an ideal Fermi--Dirac~(FD) gas, the grand
partition function reads~\cite{pathria1993,huang1987}
\be
Z_1 =
\tr\,\left( e^{-\beta\hat H^*} \right) = \sum_{N=0}^{\infty} \,
{\sum_{\{n_k\}}}' \, e^{-\beta\sum_{k}\epsilon_k^* \, n_k}
\ee
where the $n_k$'s take just two values (0 or 1) and the primed
summation means that they add up to $N$. This fermionic partition
function can be expanded as
\be
Z_1 = 1 + \sum_{N=1}^{\infty} \;
\sum_{k_1<\ldots<k_N=1}^{\infty} e^{-\beta\sum_{i=1}^{N}
\epsilon_{k_i}^*}
\label{Z11}
\ee
in a form that, on the one hand, emphasizes the contribution of
each possible value of $N$ and, on the other, illustrates the
manner of performing the trace operation. Note that, if the number
of s.p.\ levels is finite, say $K$, both infinite sums in
Eq.~(\ref{Z11}) nicely terminate when one reaches  $k_i=K$ and
$N=K$.  This way of performing the pertinent sums, which is not
the typical textbook procedure to deal with thermodynamic
quantities in the case of Fermi systems, is of a general quantal
({\it fermionic}) character, no matter what the summands' content
is. One could have, for instance
\be
\tr\,\left[\varphi(-\beta \hat H^*)\right]
= 1 + \sum_{N=1}^{\infty} \;
\sum_{k_1<\ldots<k_N=1}^{\infty} \varphi\left(-\beta \,\,\sum_{i=1}^{N}
\epsilon_{k_i}^*\right),
\label{quefuerte}
\ee
involving an arbitrary analytical function $\varphi$. If $\varphi$
is a generalized $q$-exponential we obtain Eq.~(\ref{Zqfer})
below. Indeed, in the framework of the $q$-thermostatistics one
deals with the same sort of expansion, with a crucial difference:
{\it one faces, instead of the trace of ordinary exponentials, the
trace of $q$-exponentials}. Without recourse to the integral
transform methods discussed in the Appendix, one can easily {\it
circumvent the main problem of the generalized Tsallis'
treatment}: the fact that $q$-exponentials do not follow the
distributive law with respect to sums over states, an important
result of the present endeavor.

Using the above considerations one thus recasts the Tsallis trace
in Eq.~(\ref{Zqbarg}) in the fashion
\be
\bar{Z}_q =
\left(\frac{\beta}{\beta_q}\right)^{1/(1-q)} \left[1 + \sum_{N\geq
1} \; \sum_{k_1<\ldots<k_N} \left(1-(1-q)\beta_q\sum_{i=1}^N
\epsilon_{k_i}^*\right)^{1/(1-q)}\right]
\label{Zqfer}
\ee
where, however, one has to take proper account of the cutoff
requirement. Let us analyze this important point with some detail.

The cutoff condition can be stated in the following manner
\bea
q<1: & \qquad & \sum_{i=1}^N \epsilon_{k_i}^* <
\frac{1}{(1-q)\beta}+U_q^* \label{cutoffql1} \\
q>1: & \qquad & \sum_{i=1}^N \epsilon_{k_i}^* >
\frac{1}{(1-q)\beta}+U_q^*
\label{cutoffqg1}
\eea
for every possible (ordered) configuration $(k_1,\ldots,k_N)$, and
for all $N=1,\ldots,K$. In other words, those configurations not
fulfilling the above inequalities {\it do not} contribute to the
trace, i.e., the concomitant configurational characteristic is set
to zero. Notice that the right side of both inequalities is
nothing but $[(1-q)\beta_q]^{-1}$ and, in principle, could have
positive or negative sign. Nevertheless, it can be easily seen
that the requirement of positive probabilities --cutoff
condition-- is straightforwardly fulfilled by negative-definite
``displaced" hamiltonians (whose spectrum is $\epsilon_k-\mu$) in
the case $q<1$ with $U_q^*>-[(1-q)\beta]^{-1}$, and by
positive-definite ones for $q>1$ with $U_q^*<[(q-1)\beta]^{-1}$.
In these two cases one can be sure that {\it all} configurations
contribute with non-zero probability; but any other situation
should be handled with some care. It is interesting to notice that
the same sort of conditions are found in Ref.~\cite{se_0302094} in
the context of integral transform methods for the un-normalized
(second flavor), canonical-ensemble problem. However, let us
stress that in the present situation the analytic calculations can
be fully implemented, with no other hardship than properly
accounting for Tsallis cutoff as already
mentioned, 
in  summing over states in $\bar Z_q$ (and other thermodynamic
quantities).

Let us now introduce, for the sake of brevity, the auxiliary
scalar quantity
\be
g_q(k_1,\ldots,k_N) = 1-(1-q)\beta_q\sum_{i=1}^{N}\epsilon_{k_i}^*
\ee which corresponds to the eigenvalue of the operator
$\hat{g}_q$ for the state with $N$ occupied levels:
$n_{k_1}=\ldots=n_{k_N}=1$, otherwise $n_k=0$. In terms of these
new scalars the partition function reads \be \bar{Z}_q =
\left(\frac{\beta}{\beta_q}\right)^{1/(1-q)} \left[1+ \sum_{N\geq
1} \; \sum_{k_1<\ldots<k_N} g_q(k_1,\ldots,k_N)^{\,
1/(1-q)}\right] \label{Zqfer1} \ee Moreover, one can obtain
$U_q^*$ from Eq.~(\ref{generg}), evaluating the traces as
exemplified with reference to $\bar{Z}_q$. Thus
\be U_q^* =
\frac{ \sum_{N\geq 1} \; \sum_{k_1<\ldots<k_N}\left(\sum_{i=1}^N
\epsilon_{k_i}^*\right) g_q(k_1,\ldots,k_N)^{\, q/(1-q)}} {1+
\sum_{N\geq 1} \; \sum_{k_1<\ldots<k_N} g_q(k_1,\ldots,k_N)^{\,
q/(1-q)}} \ee
Again, $U_q$ and $N_q$ will also be given by similar
expressions, replacing $\epsilon_{k_i}^*$ by either
$\epsilon_{k_i}$ or $-\mu$, respectively. One easily ascertains
that $U_q^*$ becomes $U_1^*=U_1-\mu N$ in the extensive limit, and
$U_1^*=-\partial \ln Z_1/\partial\beta$.  Let us comment that in
the case one wishes to determine, for instance, the dependence of
the $q$-energy with temperature, one should {\it consistently}
work out the above expressions in order to solve for the desired
thermodynamic quantity.

We tackle finally the evaluation of the $q$-mean value of the
occupation number operator, $\langle\hat n_l\rangle_q$. To such an
effect we, again, evaluate the traces in the form indicated above
and obtain an {\it exact expression} for the generalized fermion
occupation numbers \be \langle\hat n_l\rangle_q =
\frac{\sum_{N\geq 1} \; \sum_{k_1<\ldots<k_N} \left(\sum_{i=1}^N
\delta_{l k_i}\right) g_q(k_1,\ldots,k_N)^{\; q/(1-q)}}
{1+\sum_{N\geq 1} \; \sum_{k_1<\ldots<k_N} g_q(k_1,\ldots,k_N)^{\;
q/(1-q)}} \label{nqfer} \ee This expression can be further worked
out employing a property of the auxiliary quantities
$g_q(k_1,\ldots,k_N)$, which are not altered under a permutation
of indices. Thus, we obtain \be \langle\hat n_l\rangle_q =
\frac{g_q(l)^{\; q/(1-q)} + \sum_{N\geq 2} \;
\sum_{k_1<\ldots<k_{N-1}(k_i\neq l)} \;
g_q(k_1,\ldots,k_{N-1},l)^{\; q/(1-q)}} {1+\sum_{N\geq 1} \;
\sum_{k_1<\ldots<k_N} g_q(k_1,\ldots,k_N)^{\; q/(1-q)}} \ee Notice
that $\langle\hat n_l\rangle_q$ depends on the generalized
internal energy $U_q^*$ through $\beta_q$.

\section{Approximate results for $q\sim 1$:
comparison with previous work}
\label{SECTqapprox1}

Since the exact results discussed above exhibit a rather
formidable appearance, and in order to gain a better grasp of the
nonextensive thermostatistics of the Fermi gas, it is useful to
consider the situation $q \rightarrow 1$. It is obligatory in this
context to cite the pioneer work of B\"{u}y\"{u}kk{\i}l{\i}\c{c}
{\it et al.}~\cite{bdg_pla197}, in which quantum gases are tackled in
approximate fashion. They evaluate the partition function by
recourse to the so-called Factorization Approach~(FA), whose
essential feature is that of
ignoring interparticle correlations for the case of a dilute
quantum gas. In other words, this is equivalent to treat the
$q$-exponentials~(\ref{expq}) as if they were {\it ordinary}
exponentials, what is approximately true when $q\sim 1$.
The ensuing average occupation numbers have been
widely employed in the literature~\cite{tsallisURL}. It is to be
stressed that these FA results were developed for the unnormalized
second Tsallis-flavor mentioned in the introduction, namely, the
Curado--Tsallis formulation (recently, some of us have brought
this approximation up to date using the OLM recipes and applied
the ensuing results to the black-body problem~\cite{cuerponegro}).

Since we are here in possession of exact results for the Fermi gas, we can
indeed perform a check on the accuracy of the Factorization
Approach. If we treat the $q$-exponentials as if they were
ordinary exponentials we have
\[
g_q(k_1,\ldots,k_N)^{\; 1/(1-q)}
\approx \prod_{i=1}^N g_q(k_i)^{\; 1/(1-q)}
\]
Strictly speaking, we are making use of an approximation for the
$q$-exponential of a sum of entities in the form
\[
\expq\left(\sum_{i=1}^N x_i\right) \approx \prod_{i=1}^N \expq(x_i)
\]
which is valid for $q$ sufficiently close to 1 such that the
following inequality holds: \ $\left|(1-q)\left[\left(\sum
x_i\right)^2- \sum x_i^2\right]\right|\ll 1$. In our case,
$x_i=-\beta_q\,\epsilon_{k_i}^*$.

Therefore we obtain, from Eq.~(\ref{nqfer}), the following
FA-inspired approximate expression
\be
\langle\hat
n_l\rangle_q\approx \frac{\sum_{N\geq 1} \; \sum_{k_1<\ldots< k_N}
\left(\sum_{i'=1}^N \delta_{l k_{i'}}\right) \; \prod_{i=1}^N
\left[1-(1-q)\beta_q\,\epsilon_{k_i}^*\right]^{q/(1-q)}}
{1+\sum_{N\geq 1} \; \sum_{k_1<\ldots<k_N} \; \prod_{i=1}^N
\left[1-(1-q)\beta_q \, \epsilon_{k_i}^*\right]^{q/(1-q)}} \ee
that can be cast as \be \langle\hat n_l\rangle_q\approx
\frac{\left[1-(1-q)\beta_q \, \epsilon_{l}^*\right]^{q/(1-q)} \;
\prod_{k,k\neq l} \left\{1+\left[1-(1-q)\beta_q \,
\epsilon_{k}^*\right]^{q/(1-q)}\right\}} {\prod_{k}
\left\{1+\left[1-(1-q)\beta_q \,
\epsilon_{k}^*\right]^{q/(1-q)}\right\}} \ee leading
straightforwardly to \be \langle\hat n_l\rangle_q\approx \frac{1}
{1+\left[1-(1-q)\beta_q \, \epsilon_{l}^*\right]^{-q/(1-q)}}
\label{nfermi}
\ee
which is the Factorization Approach result of
Ref.~\cite{bdg_pla197}, except that the power $1/(1-q)$ in the result
under an unnormalized context is changed to $q/(1-q)$ under OLM
strictures. It is clear then that the FA approximation is
reasonably consistent in the $q\rightarrow 1$ limit.

By using Eq.~(\ref{nfermi}) one arrives to (formally) simple
expressions for both the number of particles and the internal
energy,
\be
N_q \approx \sum_k \frac{1} {\left[1-(1-q)\beta_q \,
\epsilon_{k}^*\right]^{-q/(1-q)}+1}
\label{Nqf}
\ee
\be
U_q
\approx \sum_k \frac{\epsilon_k} {\left[1-(1-q)\beta_q \,
\epsilon_{k}^*\right]^{-q/(1-q)}+1}
\label{Uqf}
\ee
similar to the ones obtained using the FA. We dare say that the
present treatment is simpler than the one found in
Ref.~\cite{turcos}. The simple appearance we are emphasizing here
is deceptive, though. In order to perform any practical
calculation one has to solve a coupled system due to the presence
of $\beta_q$. This problem, in turn, can be overcome by noticing
that $\beta_q$ satisfies the approximate relation
\be
\frac{\beta}{\beta_q} \approx 1+(1-q) \beta \sum_k
\frac{\epsilon_k^*} {\left[1-(1-q)\beta_q \,
\epsilon_{k}^*\right]^{-q/(1-q)}+1}
\label{bbq}
\ee
which allows one to obtain $\beta_q$ in terms of $\beta$ and, as
a consequence, to decouple Eqs.~(\ref{Nqf}) and~(\ref{Uqf}).

\hfill

In this context, which we recall is valid for $q\sim 1$, we will
now discuss the thermodynamic limit inspired in calculations
performed by Ubriaco~\cite{u_pre60}. We consider a system of
massive spinless particles in a volume $V$ at temperature $T$.
Going over to the thermodynamic limit in Eqs.~(\ref{Nqf})
and~(\ref{Uqf}) we find that
\be
N_q \approx \frac{V}{\lambda_T^3}
\left(\frac{\beta}{\beta_q}\right)^{3/2} f_{3/2}^*(z,q)
\label{NqUbriaco}
\ee
and
\be
U_q \approx \frac{3}{2} \,  T \frac{V}{\lambda_T^3}
\left(\frac{\beta}{\beta_q}\right)^{5/2} f_{5/2}^*(z,q)
\label{UqUbriaco}
\ee
where $\lambda_T=h/\sqrt{2\pi m  T}$ is the usual thermal
wavelength and $z$ is the fugacity. We have introduced the
following Fermi-like integral
\be
f_n^*(z,q) \equiv \frac{1}{\Gamma(n)}
\int_0^\infty dx \,
\frac{x^{n-1}}{[1+(q-1)x-(q-1)\frac{\beta_q}{\beta}\ln
z]^{q/(q-1)}+1}
\label{fn*zqUbriaco}
\ee
Using the definition of Eq.~(\ref{expq}), the first term in the
denominator can be re-expressed as
$\left[\expq(-x+(\beta_q/\beta)\ln z)\right]^{-q}$ so that
\be
f_n^*(z,q) =
\frac{1}{\Gamma(n)} \int_0^\infty dx \, \frac{x^{n-1}}
{\left[\expq(-x+(\beta_q/\beta)\ln z)\right]^{-q}+1}
\label{fn*zqUbriaco1}
\ee
and in this form it is easy to see that it gives the expected
result $z^{-1} e^x$ when $q=1$.

The ratio $\beta/\beta_q$ can be obtained from the thermodynamic
limit of Eq.~(\ref{bbq}) as
\be
\frac{\beta}{\beta_q} \approx
1+(1-q)\frac{V}{\lambda_T^3} \left[ \frac{3}{2}
\left(\frac{\beta}{\beta_q}\right)^{5/2} f^*_{5/2}(z,q) - \ln z
\left(\frac{\beta}{\beta_q}\right)^{3/2} f^*_{3/2}(z,q) \right]
\ee
Making use of the approximate results in Eqs.~(\ref{NqUbriaco})
and~(\ref{UqUbriaco}) we can express it in the fashion
\be
\frac{\beta}{\beta_q} \approx \frac
{1+(q-1) \, N_q \, \ln z}
{1+(q-1) \, N_q \, \frac 32 \, \frac{f_{5/2}^*(z,q)}{f_{3/2}^*(z,q)}}
\ee
Then, we can write down the generalized energy per particle
$u_q\equiv U_q/N_q$ in a useful form as
\be
u_q \approx \frac 32 \, T \, \frac{f_{5/2}^*(z,q)}{f_{3/2}^*(z,q)} \,
\frac{1+(q-1) \, N_q \, \ln z}{1+(q-1) \, N_q \, \frac 32 \,
\frac{f_{5/2}^*(z,q)}{f_{3/2}^*(z,q)}}
\ee
that resembles the conventional result.

Let us discuss the behavior of the generalized specific heat per
particle in the present context. We have computed it following the
usual procedure, by taking the temperature derivative of $u_q$
keeping the volume as well as $N_q$ fixed. The resulting
expression --not given here-- for ${C_V}_q/N_q$ is somewhat
involved; it can be written in terms of $q-1$, $N_q$, $\ln z$ and
the Fermi-like integrals $f_n^*(z,q)$ with $n=1/2, 3/2$ and 5/2.
An interesting study is to compare this generalized result against
its conventional counterpart, $C_V/N$, to which it approaches when
$q\rightarrow 1$. We have accomplished this comparison assuming
that the value of the non-extensivity parameter $q$ was close
enough to 1 that we were allowed to use Taylor expansions for all
the generalized quantities involved, up to first order in $q-1$.
We therefore defined
\be
{C_V}_q/N_q \equiv C_V/N \, \left[ 1 + (q-1) \, {\cal C}^{(1)} +
{\cal O}\left( (q-1)^2 \right) \right]
\ee
and obtained the relative first-order correction ${\cal C}^{(1)}$
as a (rather complicated) function of $z$ and $N$. In order to see
the effects of non-extensivity on the specific heat for an ideal
Fermi gas, we have considered the two extreme regimes of very low
and very high temperatures. Our main conclusions are that: \
(i)~${C_V}_q(T=0)=0$ for arbitrary $q$; \ (ii)~when $T\gtrsim 0$
(in which case $\ln z$ is of the order of $\mu_F/T \gg 1$, where
$\mu_F$ stands for the Fermi energy), ${\cal C}^{(1)}$ represents
a {\it positive} contribution that behaves as $N \, \ln z$ plus
smaller terms, then ${C_V}_q/N_q\lessgtr C_V/N$ for $q\lessgtr 1$;
and \ (iii)~when $T\rightarrow\infty$ (in which case $z$ is
approximately $\lambda_T^3 \, N/V \ll 1$), ${\cal C}^{(1)}$
represents a {\it negative} contribution that also behaves mainly
as $N \, \ln z$, then ${C_V}_q/N_q\gtrless C_V/N$ for $q\lessgtr
1$ --in the case $q\gtrsim 1$, the same finding is reported by
Ubriaco~\cite{u_pre60}--.

Finally, let us point out the differences between the present
results and the calculations of Ref.~\cite{u_pre60}. They manifest
in the presence of powers of $(\beta/\beta_q)$ in
Eqs.~(\ref{NqUbriaco}), (\ref{UqUbriaco}), and
(\ref{fn*zqUbriaco}). These differences are ultimately due to the
definition one chooses for the generalized mean values. As
discussed in the previous sections, instead of using unnormalized
mean values we work here within a {\it normalized} Tsallis
framework, in its OLM version. Moreover, we compute, for
consistency, the $q$-generalized expectation value for the number
operator instead of dealing simply with $\langle \hat N\rangle_1$.

\section{The thermodynamic limit}
\label{SECTthermodlimit}

Let us discuss the thermodynamic limit in a more general context,
and present our results for the internal energy and specific heat
of an ideal fermion gas in the thermodynamic limit for any
$q>1$-value (in the $0<q<1$ case we encounter a serious
convergence problem on account of the cutoff condition. It is only
easily tractable in the $q\rightarrow 1$ case). Following usual
practice, in the limit of large volume (coordinate space) we can
convert summations over discrete single-particle levels into
integrations in phase space. If the energy spectrum of a particle
in the gas is of the form $\epsilon({\bf p})=A\,|{\bf p}|^s$ with
degeneracy $g$, we can write these integrals in terms continuous
single-particle energies. In doing so we need the density of
states, given by
\be
D(\epsilon) \, d\epsilon = g \, \frac{L^d}{h^d} \,
\frac{2\,\pi^{d/2}}{\Gamma(d/2)}\, \frac{1}{A\,s}\, \left(
\frac{\epsilon}{A} \right)^{d/s-1} d\epsilon \equiv a \,
\epsilon^{b-1} \, d\epsilon
\label{dens_estados}
\ee
where we have assumed that the gas is contained in a hypercube of
volume $L^d$ in a $d$-dimensional space. In the case of massive
spinless particles in 3D, one has $a=2\pi V (2m)^{3/2}/h^3$ and
$b=3/2$; while for electrons in the ultra-relativistic limit one
has $a=8\pi V/(hc)^3$ and $b=3$. The conventional ($q=1$) grand
partition function for an ideal fermion gas whose density of
states is of the form~(\ref{dens_estados}), is given
by~\cite{pathria1993,huang1987}
\begin{equation}
\ln Z_1(\{\beta,\alpha\},V\rightarrow\infty) = a\,\Gamma(b)\,
\frac{f_{b+1}(e^{-\alpha})}{\beta^b}
\label{Z1}
\end{equation}
with $\beta>0$ and $b>0$. The function $f_n$ stands for the usual
FD integral, which in terms of the fugacity $z=e^{-\alpha}$ reads
\[
f_n(z) = \frac{1}{\Gamma(n)} \int_0^\infty dx \, \frac{x^{n-1}}
{z^{-1}\,e^x + 1} = \sum_{l=1}^\infty (-1)^{l-1} \frac{z^l}{l^n}
\]

One can now write down the generalized grand partition function
$\bar Z_q$ making use of an integral representation (see
Appendix). We consider here the real representation (Hilhorst
transform) of Eq.~(\ref{ZqbarHilhorst}). To this end, the function
$Z_1$ is evaluated for the transformed Lagrange multipliers
$\beta'=t(q-1)\beta$ and $\alpha'=t(q-1)\alpha$. Thus, in the
thermodynamic limit we find, for any index $q$ greater than one
(and provided that the grand-canonical configurational
characteristic is positive definite),
the following exact expression
\be
\bar{Z}_q = \frac{1}{\Gamma\left(\qunoinv\right)}
\int_0^\infty dt \, t^{\qunoinv-1} \exp\left(-t[1-(\quno)I_q]
+a\,\Gamma(b)\,
\frac{f_{b+1}(e^{-t(q-1)\alpha})}{[t(q-1)\beta]^b} \, \right)
\label{Zqbartdlim}
\ee
where $I_q$ stands for $\beta U_q+\alpha N_q=\beta U_q^*$, from
which one gets $1-(q-1)I_q=\beta/\beta_q$.

\subsection{Classical limit}

Let us consider now the case of sufficiently small values of $z$,
which corresponds to the classical limit. In such a case, the FD
integrals that appear in the different thermodynamic quantities of
interest can be expanded in the fugacity, up to second order, as
$f_n(z)\simeq z-z^2/2^n$. The ensuing OLM generalized partition
function becomes then
\bea
\bar Z_q = \expq(I_q) & + & a \,
\frac{\Gamma(b)\,\Gamma\left(\qunoinv-b\right)}{\Gamma\left(\qunoinv\right)}
\,\expq(I_q-\alpha)\left(\frac{1-(q-1)(I_q-\alpha)}{(q-1)\beta}\right)^b
\nonumber \\
&-& \frac a2 \,
\frac{\Gamma(b)\,\Gamma\left(\qunoinv-b\right)}{\Gamma\left(\qunoinv\right)}
\,\expq(I_q-2\alpha)\left(\frac{1-(q-1)(I_q-2\alpha)}{2(q-1)\beta}\right)^b
\nonumber \\
&+& a^2 \,
\frac{\Gamma(b)^2\,\Gamma\left(\qunoinv-2b\right)}{\Gamma\left(\qunoinv\right)}
\,\expq(I_q-2\alpha)\left(\frac{1-(q-1)(I_q-2\alpha)}{(q-1)\beta}\right)^{2b}
+\ldots
\eea
In the process we must take care of some restrictions on the
parameters that arise from the use of the definition of the gamma
function Eq.~(\ref{gammaR}). They are: i) $q>1$, ii)
$1-(q-1)I_q>0$, and iii) $1/(q-1)-2b>0$. As a consequence, the
present findings are valid within the region
$1<q<\min\{1+1/(2b),1+1/I_q\}$. We can also obtain $U_q$ and $N_q$
in this case, in the form prescribed in Eq.~(\ref{nevHilhorst}).
Keeping only terms up to first order in the fugacity, we are able
to express the ratio between the generalized mean energy and
particle number in the fashion
\be
u_q \equiv \frac{U_q}{N_q} \simeq b \,  T \, \frac{1-(q-1)[\beta
U_q+\alpha (N_q-1)]}{1-(q-1)b} \ee from which we can obtain
\be
u_q \simeq b \,  T \, \frac{1-(q-1)\,\alpha
(N_q-1)}{1+(q-1)\,b(N_q-1)} \ee where one can easily recognize the
appropriate result for the classical energy per particle when
$q=1$. In the present situation we also find
$I_q=(b+\alpha)N_q/[1+(q-1)b(N_q-1)]$. The implicit relation
between $\langle\hat N\rangle_q$ and its corresponding Lagrange
multiplier can be cast in the following way
\bea
&& N_q \left( \frac{1-(q-1)\,\alpha(N_q-1)}{1-(q-1)\,(\alpha
N_q+b)} \right)^{\qunoinv+1} \left(
\frac{1+(q-1)\,b(N_q-1)}{1-(q-1)\,\alpha(N_q-1)} \right)^b
\nonumber \\
&& \qquad = a \Gamma(b) \,
\frac{\Gamma\left(\qunoinv-b\right)}{\Gamma\left(\qunoinv\right)}
\, \expq(b) \, \left(\frac{\qunoinv-b}{\beta}\right)^b
\eea
from which one could, in principle, solve for $\alpha$ as a
function of $N_q$. Introducing this result into the expression
given above for $u_q$, one finally would obtain the generalized
mean energy per particle as a function of temperature (and of
$N_q$). Therefore, in this limit we are able to overcome the
 eventual possible complications
introduced, within the OLM formalism, by the presence of the term
$I_q=\sum_{j=1}^M\lambda_j\,\langle\hat O_j\rangle_q$ in the
expressions for the generalized partition function and other
thermodynamic quantities.

\subsection{Low temperature limit}

Let us face now the particular situation in which the fugacity
is extremely large:
\[
z\gg 1 \qquad {\rm i.e.} \qquad \alpha\rightarrow -\infty
\]
which corresponds to the low temperature limit with finite
chemical potential. Notice that, in this case, the transformed
variable $e^{-t(q-1)\alpha}$ becomes very large as well. We can
make use of Sommerfeld's lemma~\cite{s_zp47,pathria1993}
\be
f_n(z) = \frac{(\ln z)^n}{\Gamma(n+1)}
\left[1+\frac{\pi^2}{6}n(n-1)\frac{1}{(\ln z)^2}+
{\mathcal{O}}\left(\frac{1}{(\ln z)^4}\right)\right]
\ee
which gives an asymptotic expansion of the FD integrals for $z\gg
1$. As can be seen, the dominant term in the expansion is of order
$(-\alpha)^n$. We can evaluate the integral in
Eq.~(\ref{Zqbartdlim}) up to this order, getting an expression
valid for all $q>1$ in the low temperature limit:
\be
\bar{Z}_q \simeq
\expq\left(I_q-\frac{a}{b(b+1)} \, \left(\frac{-\alpha}{\beta}\right)^b
(-\alpha) \right)
\ee
Additional quantities evaluated up to the same order are
\be
U_q \simeq \frac{a}{b+1}\,\mu^{b+1} \qquad \textrm{ and } \qquad
N_q \simeq \frac{a}{b}\,\mu^b \ee independently of the value of
$q$, in agreement with previous results obtained using the OLM
formalism~\cite{OLM}. Using the above relations we can derive the
Fermi potential as $\mu_F=(N_q \, b/a)^{1/b}$; and we can also
identify the Fermi temperature as $T_F\equiv\mu_F=(N_q \,
b/a)^{1/b}$. Besides, it is easy to obtain $I_q^{(0)}$
and then simplify the expression for $\bar Z_q$, leading to
\be
\bar{Z}_q \simeq \expq\left( 2 \, I_q^{(0)}\right) =
\expq\left(-\frac{2 a}{b(b+1)} \, \beta\mu^{b+1} \right)
\ee
It is worth pointing out that the mean energy per particle in the
OLM-Tsallis framework {\it does not depend explicitly on the value
of $q$}. Indeed, in the limit under consideration we have
$U_q^{(0)}/N_q=\mu_F \, b/(b+1)$.

Performing calculations up to the next order of approximation we
arrive, after a little algebra, at the following results which are
valid for any $q>1$:
\be
N_q \simeq \frac{a}{b}\,\mu^b \, \left[ 1+\frac{\pi^2}{6} \,
(b-1)b \left( \frac{T}{\mu} \right)^2 \frac{{\mathcal I}^{\,
2}}{1-(\quno)} + \ldots \right] \ee where the quantity ${\mathcal
I}\equiv
1-(\quno)\beta\{U_q-\mu[N_q-a\,\mu^b/b(b+1)]\}=1-(q-1)(I_q-I_q^{(0)})$
is very close to unity. Actually, the second term in ${\mathcal
I}$ vanishes if one keeps just terms corresponding to the
lowest-order approximation, which eliminates the term $I_q$
introduced by the OLM procedure (see Eqs.~(\ref{constraints}),
(\ref{charact}), and~(\ref{charactGC})~). For the chemical
potential we find
\be
\mu^{(2)} = \mu_F \left[ 1-\frac{\pi^2}{6} \, (b-1) \left( \frac{
T}
{T_F}
\right)^2 \frac{{\mathcal I}^{\, 2}}{1-(\quno)} \right]
\ee
whereas the generalized mean energy becomes
\be
U_q^{(2)} = \frac{a}{b+1}\,\mu_F^{\ b+1} \, \left[
1+\frac{\pi^2}{6} \, (b+1) \, \left( \frac{T}
{T_F}
\right)^2
\frac{{\mathcal I}^{\, 2}}{1-(\quno)} \right]
\ee
{}From this expression we derive the generalized specific heat at
constant volume
\be
{C_V}_q^{(2)} =
N_q  \, \frac{\pi^2}{3} \, b \, \frac{ T}
{T_F}
\,
\frac{1}{1-(\quno)} \ee where we have assumed ${\mathcal I}^{\,
2}=1$. Notice the linear behavior of the specific heat with
temperature and that the only difference with the conventional
results (see, e.g., Ref.~\cite{pathria1993}) comes through a
factor $1/(2-q)$ which goes to unity in the limit $q\rightarrow
1^+$.

\section{Conclusions}
\label{SECTconclusions}

In this communication we have presented an exact statistical
treatment for the ideal Fermi system in the generalized,
nonextensive thermostatistics framework of the third
Tsallis-flavor, the TMP one. Our main innovation is that of
employing the OLM approach to
nonextensivity~\cite{OLM,virial,gasideal,temperature,ley0,cuerponegro},
which allows one to obtain analytical results unavailable if one
uses other algorithms that revolve around the concept of Tsallis'
entropy.

Additionally, we have i)~solved in exact fashion the Fermi-TMP
equations, ii)~introduced a method for evaluating traces that
bypasses the use of the Gamma representation, iii)~devised a
rather simple treatment of $q \approx 1$ instances, and
iv)~studied interesting limiting situations. More specifically,
the exact generalized partition function in the grand canonical
ensemble has been given, and we derived the ensuing statistics for
arbitrary positive values of the nonextensivity index $q$. Several
limit instances of interest were here discussed in some detail:
\begin{enumerate}
\item the thermodynamic limit,
\item the case $q\sim 1$ ($q=1$ corresponds to the conventional
Fermi--Dirac statistics),
\item the low temperature regime, where we obtained results that
are independent of the specific $q$-values.
\end{enumerate}
In writing down the generalized expectation value for the
occupation number operator we were able to explicitly display the
{\it correlation} among the occupations of different levels, which
is a typical nonextensive effect. Indeed, the distinct mean
occupation numbers are seen to disentangle from each other as one
approaches the conventional $q=1$ statistics. Finally, we
discussed the limits of validity of the Factorization Approach of
B\"{u}y\"{u}kk{\i}l{\i}\c{c} {\it et al.}~\cite{bdg_pla197}.

\acknowledgments

The authors acknowledge financial support from CONICET
(Argentina). M.P.\ acknowledges also a grant from Fundaci\'on
Antorchas (Argentina).


\appendix{\section{INTEGRAL REPRESENTATIONS}}
\label{APintegralrepresentations}

In this appendix we summarize a practical method for calculating
the generalized partition function~(\ref{Zqbar})
and expectation values~(\ref{generf}) of relevant observables,
by recourse to integral representations based on the definition
of the Euler gamma function. We start with the identity
\be
\int_0^\infty dt \ t^{\nu-1} \, e^{-t \eta} =
\eta^{-\nu} \, \Gamma(\nu)
\label{gammaR}
\ee
for $\Re(\eta)>0$ and $\Re(\nu)>0$ (see for instance
Ref.~\cite{gr2000}, page 342). One can then write
\be
{\hat f_q}^{\; 1/(1-q)} =
\frac{1}{\Gamma\left(\frac{1}{\quno} \right)}
\int_0^\infty dt \ t^{\frac{1}{\quno}-1} \, e^{-t \hat f_q}
\ee
with the restrictions that i) $\hat f_q$ be positive (which is
indeed always complied with because of Tsallis' cutoff
condition~\cite{pp_pla193}), and ii) $q>1$. The usefulness of the
transformation becomes evident, as the power-law form is converted
into an exponential factor.

Evaluating the trace of the above expression --see final comments
below-- and introducing the OLM quantal configurational
characteristic, one arrives at an integral representation
(Hilhorst transform~\cite{tsallis1994}) for the OLM generalized
partition function (\ref{Zqbar}) of index $q>1$, in the form
\be
\bar{Z}_q =
\frac{1}{\Gamma\left(\frac{1}{\quno} \right)} \int_0^\infty dt
\ t^{\frac{1}{\quno}-1} \, e^{-t [1+(1-q)\sum_k \lambda_k \langle \hat
O_k\rangle_q]} \, Z_1(\{\lambda'_i\})
\label{ZqbarHilhorst}
\ee
Here the integrand contains the corresponding {\it conventional partition
function} $Z_1$ evaluated for the set of {\it transformed Lagrange
multipliers} $\{\lambda'_i(t)=t(q-1)\lambda_i, \, i=1,\ldots,M\}$.
Notice that this transformation preserves the sign of the Lagrange
parameters (this fact is used in the text).
The generalized expectation values~(\ref{gener}) can be
expressed, for $q>1$, in terms of integrals involving $Z_1$ and
its derivative with respect to the associated Lagrange multiplier.
Indeed, one realizes that
\be
\langle \hat
O_j\rangle_q = - \, \frac{\int_0^\infty dt \ t^{\frac{1}{\quno}} \,
e^{-t [1+(1-q)\sum_k \lambda_k \langle \hat O_k\rangle_q]}
\ \partial Z_1(\{\lambda'_i\}) / \partial\lambda'_j }
{\int_0^\infty dt \ t^{\frac{1}{\quno}} \, e^{-t
[1+(1-q)\sum_k \lambda_k \langle \hat O_k\rangle_q]} \,
Z_1(\{\lambda'_i\})}
\label{nevHilhorst}
\ee

Let us mention that one can obtain alternative integral
representations for $\bar Z_q$ and $\langle\hat O_j\rangle_q$ in
the range $0<q<1$ by recourse, for instance, to the following
complex representation of the Euler gamma function
\be
\int_{-\infty}^\infty dt \ (\zeta+i t)^{-\nu} \, e^{(\zeta+i t) \eta}
= 2\pi \, \eta^{\nu-1}/{\Gamma(\nu)}
\label{gammaC}
\ee
for $\eta>0$, $\Re(\nu)>0$ and $\Re(\zeta)>0$ (\cite{gr2000},
p.~343). In this case (that may be called the Prato--Lenzi
transform~\cite{prato,lenzi}) the partition function can be
written as
\begin{equation}
\bar{Z}_{q} =
\frac{1}{2\pi}\,\Gamma \left( \frac{q-2}{q-1}\right)
\int_{-\infty }^{\infty} dt \ (1+it)^{\frac{1}{q-1}-1} \, e^{(1+it)[1+(1-q)
\sum_k \lambda_k\langle\hat{O}_k\rangle_q]} \, Z_{1}(\{\tilde{\lambda}_i\})
\label{ZqbarPratoLenzi}
\end{equation}
where $\tilde{\lambda}_i(t)\equiv(1+it)(1-q)\lambda_i$ for each
$i=1,\ldots,M$, and the region of validity is $q<1$ or $q>2$. The
associated mean values take the form
\be
\langle \hat O_j\rangle_q = - \,\frac{
\int_{-\infty}^\infty
dt \ (1+it)^{\frac{1}{\quno}} \, e^{(1+it)
[1+(1-q)\sum_k \lambda_k \langle \hat O_k\rangle_q]} \,
\ \partial Z_1(\{\tilde\lambda_i\}) / \partial\tilde\lambda_j }
{\int_{-\infty}^\infty
dt \ (1+it)^{\frac{1}{\quno}} \, e^{(1+it) [1+(1-q)\sum_k \lambda_k
\langle \hat O_k\rangle_q]} \, Z_1(\{\tilde\lambda_i\})}
\label{nevPratoLenzi}
\ee
for $j=1,\ldots,M$ and $q<1$.

Some additional remarks are necessary. A quite detailed analysis
concerning integral representations for the $q$-thermostatistics
can be found in the recent preprint~\cite{se_0302094} by Solis and
Esguerra, who pay special attention to practical details of the
representations discussed above. One has to make sure that {\it
all} states are contributing to the evaluation of the trace --in
the sense that there is no cutoff-- in order to get $Z_1$ on the
r.h.s.\ of Eqs.~(\ref{ZqbarHilhorst}) or~(\ref{ZqbarPratoLenzi}).
Solis and Esguerra point out that this fact has not been taken
into account in the majority of Tsallis-related works~(see, for
instance,~\cite{aslslf_pa317}). In our particular case, the
condition to be imposed in order to ensure that the Hilhorst-type
representation can be safely used can be stated in the following
terms: $\hat f_q$ as given by Eq.~(\ref{charactGC}) should be
positive definite for all states, i.e., the lowest energy
eigenvalue (or the greatest lower bound of the Hamiltonian) should
be greater than or equal to $\mu(\hat N-N_q)+U_q-1/[\beta(q-1)]$.
(Notice that the simpler requirement given in
Ref.~\cite{se_0302094}, namely that $\hat H$ be greater than
$-1/[\beta(q-1)]$, originates in the fact that a
canonical-ensemble description is performed and also that
unnormalized, Curado--Tsallis mean values are employed.)

\end{document}